\documentclass[journal,twoside]{IEEEtran}
\usepackage{cite}
\usepackage{amsmath,amssymb,amsfonts}
\usepackage{algorithmic}
\usepackage[]{graphicx}
\usepackage{multicol,multirow}
\usepackage{epsfig}

\usepackage{gensymb}
\usepackage{textcomp}
\usepackage{hyperref}
\newcommand{\eg}{\textit{e}.\textit{g}., }
\newcommand{\ie}{\textit{i}.\textit{e}.\ }

\begin{document}

\title{Fast Alternating Radial Beamforming for Speed-of-Sound Imaging\\ Based on Apparent Speckle Shifts}
\author{\IEEEauthorblockN{Dieter Schweizer\IEEEauthorrefmark{1}, Can Deniz Bezek\IEEEauthorrefmark{2}, Orcun Goksel\IEEEauthorrefmark{2}
\thanks{Funding was provided partially by the Uppsala University Medtech Science and Innovation Centre.}}\\[.5ex]
\IEEEauthorblockA{%
\IEEEauthorrefmark{1}Department of Information Technology and Electrical Engineering, ETH Zurich, Switzerland\\
\IEEEauthorrefmark{2}Department of Information Technology, Uppsala University, Sweden}}

\maketitle

\begin{abstract}
Pulse-echo speed-of-sound (SoS) imaging based on minute misalignments between consecutively acquired ultrasound images traditionally relies on images beamformed on Cartesian grids. 
Existing SoS imaging developments do not allow for real-time imaging and typically do not prioritize feasibility in conventional ultrasound systems that have limited resources and rigid processing structures.
In this work, we propose a resource-efficient approach based on radial beamforming with virtual source transmits for implementation within an on-the-fly beamformer.
We also introduce alternating transmissions with fast pair-alternating  beamforming for motion-robust displacement tracking with typical line-based beamformers.
We tested these methods comparatively on numerical simulations, tissue-mimicking phantom experiments, and in vivo data from breast lesion examinations. 
We demonstrate that the proposed radial grid beamforming approach performs comparably to a Cartesian grid approach, while allowing implementation on standard hardware for beamforming.
Our proposed sequences would allow for SoS data acquisition frame rates of more than 20\,fps in parallel to conventional B-mode imaging.
The proposed speckle-shift based radial approach with fast alternation between congruent beamforming lines is a major step towards real-time SoS imaging on standard ultrasound systems with moderate resources.
\end{abstract}


\section{Introduction}
\IEEEPARstart{P}{ulse-echo ultrasound} (US) imaging is widely used in health diagnostics and therapy thanks to the non-ionizing nature, relatively low cost and compact realization of the required technology. Several modalities such as B-mode echo imaging, Doppler-based flow measurements, and more recent technologies for tissue characterization through elastography and speed-of-sound (SoS) imaging allow us to examine tissues from complementary points of view.

The propagation speed of sound waves inside a material depends on two properties, density and bulk modulus \cite{Cobold_foundations_ultrasound_2006}. Variations in either property due to changes in material composition (\eg increase in fat in the liver~\cite{Fetzer_fatty_liver_2023}, muscle degeneration due to aging~\cite{sanabria_speed_2018_sarc}, variations in breast density~\cite{Bezek_BreastDensity_2025} or pathological changes due to cancer~\cite{Schweizer_SoSasPotentialBiomarker_2025}) may hence lead to detectable SoS changes.
Research on SoS aims either to estimate a global value for the entire tissue under investigation
\cite{Ophir_Estimation_86,Anderson_direct_98,Krucker_sound_04,Napolitano_sound_06,Shin_estimation_10,Qu_average_12,Yoon_in-vitro_11,Park_MeanSoS_Quality_2011,Hasegawa_Coherence_2019,Bezek_global_22_ultrasonics}
or to generate a map (image) of local SoS distribution
\cite{jaeger_computed_2015,Imbault_2017,sanabria_spatial_2018,Jakovljevic_modelbased_SoS_2018,Bendjador_ultrafast_sos_2021, Heriard_refraction-basedSoS_2023,bezekTMI_2025}.

Most methods for imaging local SoS need flexible programmable hardware, involved processing stages, and off-line post-processing for operation.  
A fast acquisition sequence for pulse-echo SoS imaging was proposed in~\cite{Schweizer_RobustVS}.
A deep-learned loop-unrolling-based inverse-problem solver for SoS-based problems was shown in~\cite{vishnevskiy2018im,vishnevskiy_deep_2019} to operate on the order of tens of milliseconds per SoS frame.
Combining these approaches -- even considering potential overheads -- can achieve sub-second frame updates in pulse-echo SoS.
However, this requires experimental research US systems consisting of multichannel radio-frequency (RF) data acquisition and transfer to a separate computer for conducting full-frame beamforming for each wide-beam transmit. 
In order to enable clinical realization, methods compatible with hardware and processing pipeline limitations of standard US machines are needed.

The processing chain for speckle-shift-based SoS imaging requires four main steps:
First, an ultrasonic wave is transmitted either as plane waves with a few degrees of separation \cite{Stahli_improved_20} or as diverging waves with their origins separated by some millimeters \cite{Rau_divergingWave_2021, Schweizer_RobustVS}.
Second, received RF data are beamformed. 
Third, a misalignment is found between pairs of beamformed images, \eg using a speckle tracking algorithm.
Finally, using a forward model of SoS imaging, the inverse problem of its image reconstruction is solved to obtain a local SoS distribution map.

Most standard US systems are limited in resources:
Typically, only a subset of transceivers can be addressed at a time during a transmit (Tx) or receive (Rx), whereas large transducer apertures may be required for SoS imaging, \eg for generating plane waves.
Although some Tx sequences, such as virtual-source (VS) diverging waves~\cite{Schweizer_RobustVS}, may overcome such a Tx limitation, the aperture limitation due to addressable transceivers is still a challenge on the Rx side.
Most standard US front-ends with FPGA/ASIC based beamformers are designed to systematically delay the Rx data in channel-wise buffers, which are summed for each consecutive beamforming point.
The delay values are either calculated on-the-fly based on a simplified delay model or obtained using precalculated tables.
Limited temporal buffer sizes, algorithmic reliance on monotonically increasing delays, and limited processing power of a few delay/sums at each Rx cycle restrict typical on-chip beamformers to only beamform points on one or very few lines originating from a transceiver (known as Rx center).
Such a structure does not allow beamforming a multitude of points at once as performed in most research-based SoS imaging solutions but can still be re-purposed for the requirements of SoS imaging as shown in this work.

SoS imaging relies on displacement tracking between beamformed RF frames from different transmits, as apparent speckle misalignments result from local SoS differences on wave travel paths through the tissue. 
To that end, the existing SoS imaging approaches perform Rx beamforming, \eg via delay-and-sum (DAS)~\cite{Perrot_DAS_21}, for each location on the Cartesian beamforming grid for each single Tx.
This allows obtaining the entire beamformed frame with a single Tx; for instance, only two transmits are sufficient to generate the entire image frames and hence the displacement tracking between them. 
Alternatively, displacement tracking can be performed batch-wise on smaller image sections, \eg beamforming lines, that are each from separate transmits. 
Such patches/lines can be selected in a way suitable for the specific beamformer constraints, \eg a single beamforming line on a conventional system with very limited resources.
Tracking can still be performed between image patches/lines from two consecutive Tx, which reduces the inter-pair motion sensitivity down to that of conventional full-frame software beamforming for tracking.
In this work, we introduce and study this line-by-line Tx-pair alternating approach.
In particular, we present it as a step towards real-time SoS imaging on conventional ultrasound systems with line-based beamformers, using a radial beamforming grid and virtual-source diverging transmissions.

%%%%%%%%%%%%%%%%%%%%%%%%%%%%%%%%%%%%%%%%%%
\section{Methods}

\subsection{Speckle-shift based SoS imaging}
For SoS imaging, beamformed RF data from a pair of different transmits are used to track speckle misalignments $\boldsymbol{\Delta d}$ in the fast axis that are due to time delay differences $\boldsymbol{\Delta\tau} = 2\boldsymbol{\Delta  d} / c_0$ occurring with the round-trip waves propagating with actual SoS $\boldsymbol{c}$ different from the assumed beamforming SoS $c_0$.
Acoustic waves passing through different tissue locations accumulate SoS deviations between the real and assumed values along their travel path.
By taking the slowness $\boldsymbol{\sigma}=1/\boldsymbol{c}$ as the inverse of SoS, the forward model of speckle-shift based SoS imaging can be written as 
\begin{equation}\label{eq:sosforwardmodel}
 \boldsymbol{\Delta\tau} =\textbf{L}(\boldsymbol{\sigma-}\sigma_0) ,
\end{equation}
where the differential path matrix $\textbf{L}$ links the relative slowness distribution $(\boldsymbol{\sigma}-\sigma_0)$ to the relative delay measurements $\boldsymbol{\Delta\tau}$. 

SoS reconstruction is then performed via solving the following inverse problem~\cite{Rau_divergingWave_2021}:
\begin{equation}\label{eq:sosreconRadial}
 \boldsymbol{\hat\sigma} = \arg \min_{\boldsymbol{\sigma}}
 \| \textbf{L}(\boldsymbol{\sigma-}\sigma_0) - \boldsymbol{\Delta\tau} \|_1  +  \lambda \|\textbf{D}\boldsymbol{\sigma} \|_1\ \ 
\end{equation}
where the regularization matrix \textbf{D} together with the weight $\lambda$ controls the amount of spatial edge-preserving regularization, essential due to the poor conditioning of the problem. 
The optimization problem is solved using a limited-memory Broyden–Fletcher–Goldfarb–Shanno (L-BFGS) algorithm. 

\subsection{Conventional Cartesian-grid beamforming}
Given that speckle shifts are to be tracked between two images beamformed independently for two Tx sequences, separated by $d_\mathrm{ch}$, a common beamforming grid for both images is desired to avoid any (error-prone) resampling in the RF domain.
A fixed Cartesian grid (C-grid), with the fast (signal-modulation) direction in the vertical axis, then lends itself as the natural and the conventional choice as a common beamforming grid for both images.

To minimize artifactual speckle-shifts due to point spread function (PSF) rotations, there are methods~\cite{Stahli_improved_20} that aim at aligning PSFs for different Tx directions, by counter-rotating the Rx directions.
However, such alignment often requires many Rx elements outside the transducer surface, and to achieve the intended Rx angles and hence centers, it requires Rx apertures to be reduced symmetrically, causing them to be much smaller than otherwise possible.
Instead, (f-number adjusted) Rx apertures centered above each beamforming point (the so-called zero-angle Rx) independently of Tx have been successfully used in the literature for SoS estimation~\cite{Bezek_global_22_ultrasonics,Schweizer_RobustVS}.
Furthermore, using a fixed Rx aperture per beamforming point between the Tx pair ensures that the echo Rx paths coincide, hence the SoS sensitivity depending only on the Tx paths, making the imaging matrix sparser for faster and potentially more robust reconstructions.

\begin{figure*}
\centering
\includegraphics[width=0.9\linewidth]{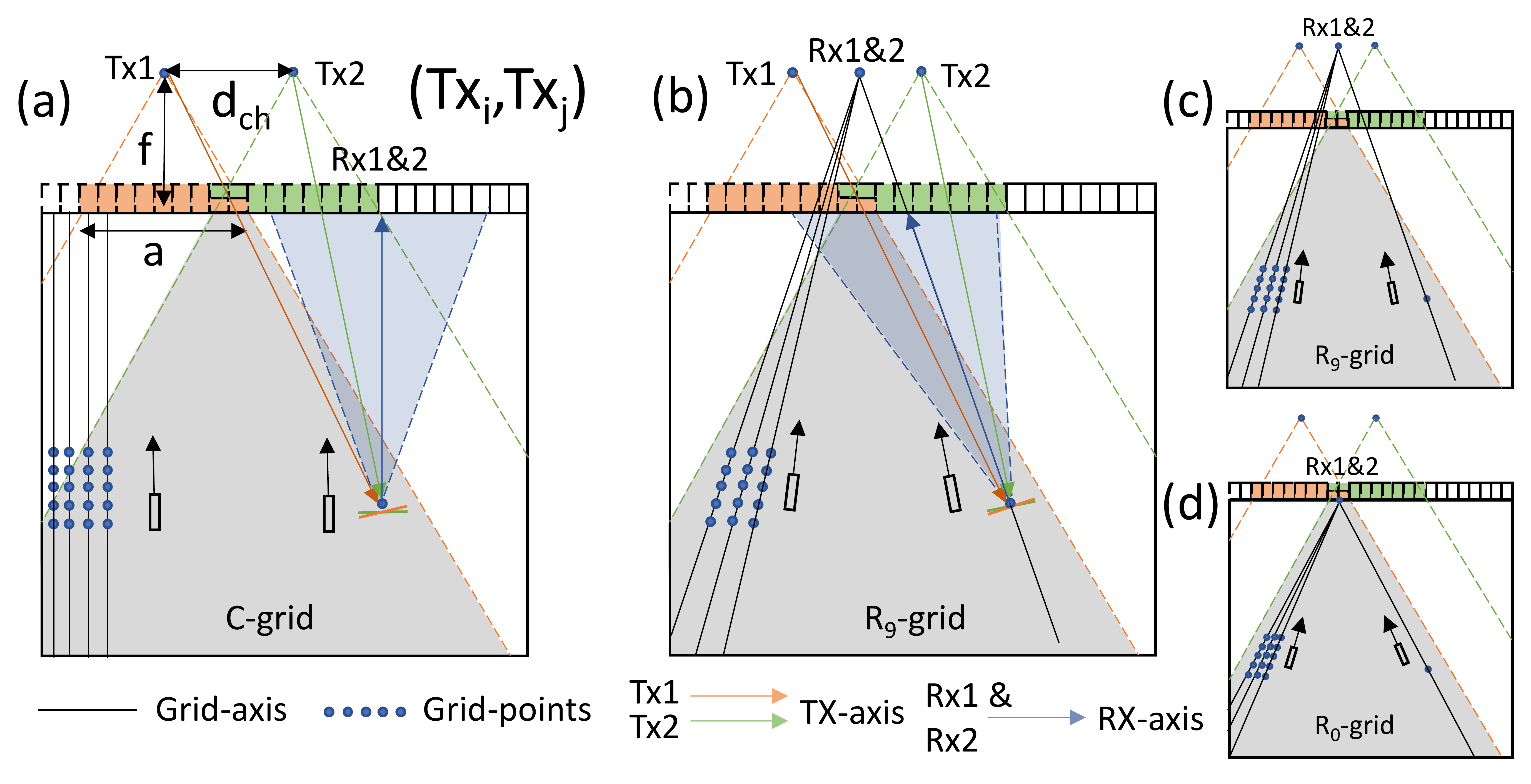} 
\caption{(a) Method of standard Cartesian-grid (C-grid) beamforming: A pair of VS transmits Tx$_i$ and Tx$_j$, separated by $d_\mathrm{ch}$ insonify the tissue. The virtual focus point is behind the transducer at a distance $f$, which results in an active transmit aperture of $a$. Rx aperture (shown in blue) for any beamforming grid point is centered above itself, with width depending on the chosen Rx f-number.
(b) In our proposed radial beamforming, the fast direction of Rx beamforming lines during both Rx1 and Rx2 are chosen to originate from an Rx origin in the middle between the respective Tx1 and Tx2 transmit focus centers.
Two variants are shown with the Rx origin (c)~on the line connecting the two Tx focus positions (R$_9$-grid) and (d)~on the transducer surface (R$_0$-grid).
The gray areas indicate the overlap of the wide beams from the Tx pair, where displacement tracking is possible as input to reconstruction. Hence, beamforming is also only required for points within this region.
The black rectangles show sample displacement tracking windows along the fast axes of beamforming.
The orange and green lines underneath each beamforming point  illustrate respective approximate PSF orientations.
}
\label{fig:BF_methods}
\end{figure*}

Figure \Ref{fig:BF_methods}(a) shows the pulse-echo wave paths for a pair of VS transmissions Tx$_i$ (orange) and Tx$_j$ (green) 
arriving at a C-grid beamforming (respectively, displacement tracking) point.
Given element directivity, the overlap of two insonifications from a Tx pair (shaded gray area in the figures) is the region that contains matching beamforming points, for which displacements can be tracked.
Given the triangular nature of VS overlaps, a Cartesian grid then includes several irrelevant points to beamform outside such overlap.
A conventional beamformer starting iteratively from the transducer surface downward would then waste valuable resources to beamform these unnecessary points.
Furthermore, for beamforming the points towards the sides of the image/transducer, part of the intended Rx aperture will fall outside the transducer surface, hence either requiring to symmetrically reduce the Rx aperture, which reduces SNR, or shifting the Rx aperture inward, which artificially rotates PSFs.
Note that such out-of-transducer Rx apertures aggravate rapidly if PSF-alignment approaches are used.
Cartesian beamforming is therefore not resource efficient for an on-the-fly beamformer.

Some low-end US systems are further limited by multiplexing capability, where they can address only the same set of transducer elements during the Tx and Rx events of a single Tx-Rx cycle.
Hence, Rx apertures that move further and hence differ substantially from the Tx elements is a further limitation of such Cartesian-grid approach.
Beyond the above acoustic and signal processing limitations, from a computational standpoint, a beamformer iterating over the fast dimension of a C-grid beamforming line will need to consider consecutive Tx distances (between vertical grid neighbors) increasing not linearly as usual in focused imaging but instead by a trigonometric or a Pythagorean function, which adds a few more operations on a resource-limited low-end or hand-held system.

\subsection{Radial Beamforming}
To overcome the above challenges, we propose an alternative approach with a radial beamforming grid (R-grid) as shown in Figure \Ref{fig:BF_methods}(b). 
R-grid points are located on Rx beam axes originating from an Rx origin. 
This is in a way similar to conventional phased-array beamforming, but with the same R-grid used across two different Tx of a pair and applied on a linear array with the Tx pairs rolling across the transducer.
To symmetrically cover the Tx overlap, the Rx origin shall be equidistant to the Tx centers, \ie lie on the vertical (axial) line in their middle.
For the axial location, one option is to place the Rx origin virtually behind the transducer similarly to the transmits.
Without loss of generality, we herein study a radial Rx origin at the same Tx focal depth (of 9\,mm in this paper), leading to a R$_9$-grid for beamforming as shown in Figure \Ref{fig:BF_methods}(c), where the suffix indicates the distance (in mm) of the virtual Rx axis origin behind the transducer surface.
This choice results in Rx apertures to be similar for different beamforming points, \ie around the intersections of the Rx-grid axis lines with the transducer surface in Figure \Ref{fig:BF_methods}(c).

Another alternative for beamforming would be an R$_0$-grid with the Rx origin on the transducer surface as in Figure \Ref{fig:BF_methods}(d).
This allows all beamforming grid points to have identical Rx aperture centers, potentially with aperture sizes changing based on the Rx f-number, which is a natural and easier to implement option on most hardware.
Furthermore, all beamformed R$_0$-grid points can be used for subsequent displacement tracking, \ie are within the overlap of the two Tx directivities, in contrast to many points of the C-grid being inapplicable.
In this paper, we compare these three beamforming grid choices, \ie the Cartesian C-grid with the radial options R$_9$-grid and R$_0$-grid.

\subsection{Radial grid in displacement tracking and SoS reconstruction}
Conventional ultrasound systems perform beamforming on-the-fly along lines by dynamically changing channel-specific delays on a set of incoming Rx echo signals. 
In phased-array imaging, such lines spread radially from an Rx origin located on or virtually behind the transducer surface.
The beamforming points then lie on a polar coordinate grid, and any subsequent displacement tracking ideally also operates along such radial lines, \ie the fast dimension of beamformed RF data. 
Displacement tracking, which is typically run as a windowed cross-correlation or a multi-point phase tracker -- indicated as black boxes in Figure \Ref{fig:BF_methods}, is therefore performed vertically with a C-grid and radially with an R-grid.
Note that the final SoS reconstruction can independently be on an arbitrary grid, including a Cartesian-grid for convenient image display, as long as the wave-path rasterization process when forming the imaging model $L$ in (\ref{eq:sosforwardmodel}) takes into account the relevant coordinate frames.

Ideally, the PSF at all beamforming locations should be identical between the Tx pairs such that any observed displacement is due solely to the medium SoS.
By having the Rx aperture centered between the existing Tx locations for all beamformed lines, the proposed radial beamforming enables much larger apertures compared to its alternatives and furthermore, R-grid paired PSFs are by design symmetric about the displacement tracking axis (see Figure \Ref{fig:BF_methods}) whereas C-grid PSFs are systematically tilted to one or the other direction, which could bias the displacement tracking process for the latter.

\subsection{Fast pair-alternating beamforming}
\label{sec:RadialTxRxSequences}
Since the imaging model associates any observed speckle shift with SoS differences, any physical motion that occurs between the acquisitions of the paired echo data for a point introduces errors in the SoS reconstruction.
In \cite{Schweizer_RobustVS}, fast acquisition sequences with minimum delay between two pairs of RF frames were shown to reduce motion-induced artifacts.
This makes the assumption that the second Tx frame should be acquired after the first one is fully completed.
Although minimizing the entire sequence timing as such is beneficial, what affects the physical motion is indeed the time elapsed between the two (paired) beamforming of each individual location.
With this in mind, we propose pair-alternating beamforming (PAB) by beamforming each grid line \emph{alternatingly} between the Tx pair, such that the beamforming time difference for a point reduces down to one Tx-Rx cycle.
This is then just a change in acquisition sequence and is possible on most beamformers regardless of hardware.

Compared to software beamforming from a single echo acquisition, line-by-line hardware beamforming of an entire image frame (\eg of $n$ lines) would typically take $n$ times slower.
To achieve high frame rates, typical beamformers are able to parallel-process to a small extent, by simultaneously beamforming multiple (\eg 2-to-8) lines from the same Rx echo. 
For focused imaging, this can then beamform a few neighboring lines within the Tx focusing width (to ensure insonification) all at once within the same Tx-Rx cycle.
As diverging waves are unfocused, we can beamform as many lines as allowed by the hardware.
We utilize this capability to beamform $\ell$ lines at a time, reducing the acquisition of a frame down by $n/\ell$; and the total for the frame pair to $2n/\ell$.
Accordingly, our method works by beamforming $\ell$ lines from Tx1, then the same $\ell$ lines from Tx2, and shifting to the next $\ell$ lines alternatingly for Tx1 and Tx2, and so on.

Note that with the PAB approach above, the paired beamformed lines can be immediately used for displacement tracking.
This obviates displacement tracking waiting for the completion of entire frame acquisition and enables the serialization and parallelization of displacement tracking together with the acquisition of subsequent lines.

\subsection{Sample implementation}
\begin{figure*}
\centering
\includegraphics[width=0.9\linewidth]{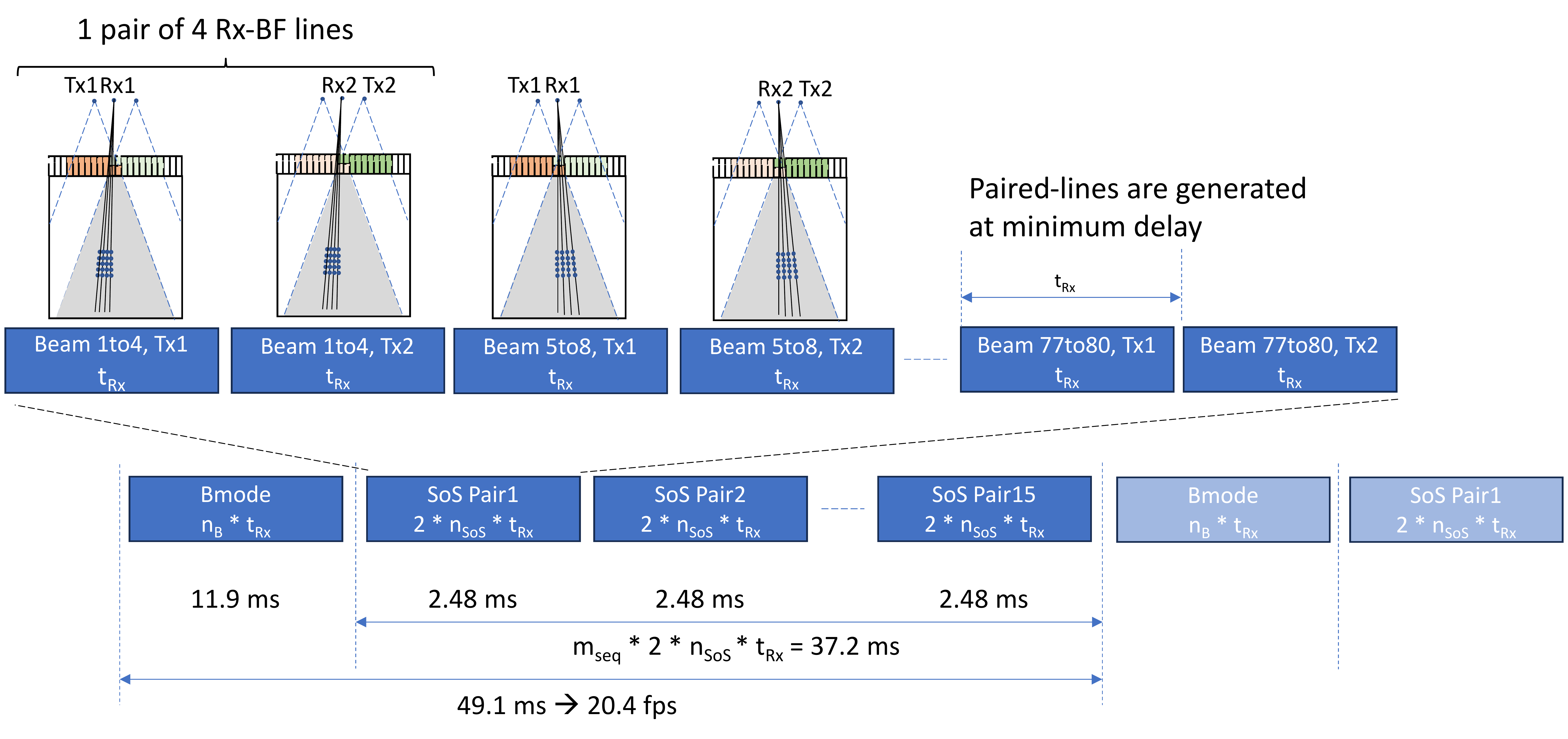} 
\caption{Tx/Rx sequence of B-mode with interleaved SoS imaging using a 4-fold parallel beamformer. The sequence is based on a fast virtual-source sequence with $m_\mathrm{seq}=15$ pairs~\cite{Schweizer_RobustVS}.
The example is given for an imaging depth $d = 40$\,mm. With an assumed slowness $\sigma_{0}=0.651\,\mu$s/mm and a beamformer reconfiguration overhead of $t_\text{o}=10\,\mu$s, the Rx-time is $t_\text{Rx} = 2d\sigma_{0}+t_\text{o}=62\,\mu$s. A typical B-mode image, with $n_\text{B}=192$ Tx-Rx cycles, can be acquired in approximately $11.9$\,ms.
An SoS sequence with a radial sector of $\pm 20 \degree$ with $0.5\degree$ angular resolution requires 80 lines, which is possible to 4-fold parallel beamform in $n_\text{SoS}=20$ Tx-Rx cycles, taking $2.48$\,ms for all the acquisition and beamforming of a pair.
Conducting this 15 times plus the B-mode timing requires $49.1$\,ms, indicating the feasibility of 20+ frames-per-second dual-mode acquisition on conventional hardware.
}
\label{fig:BF_RadialTxRxSequence}
\end{figure*}

In the following, we provide a reference implementation as an example.
Assuming a radial angular beamforming field-of-view of $\pm 20\degree$ with an angular resolution of $r=2$ beams per degree, for the Tx pair $40x2x2=160$ lines need to be beamformed in total.
With a parallel beamforming factor $\ell=4$, this then requires 40 Tx/Rx cycles, with each cycle requiring about 52\,$\mu$s for a 4\,cm depth plus an internal update time of the beamformer parameters of about 10\,$\mu$s for sequence switching.
Figure \Ref{fig:BF_RadialTxRxSequence} illustrates a detailed timing for this sample sequence
that yields the acquisition of 15 Tx-pair SoS data in 37.2\,ms, allowing for imaging frame-rates of over 20\,fps even when interleaved with B-mode imaging in parallel.
If higher B-mode frame-rates are desired and/or if SoS post-processing is not fast enough, SoS Tx-pair acquisitions can also easily be spread across multiple B-mode images.

\section{Experiments}
Experiments with numerical simulations, tissue-mimicking phantoms, and in vivo data have been conducted, evaluating the SoS images with accuracy and contrast metrics.

\subsection{Data}
Numerical simulations were performed in MATLAB (2023b, The MathWorks Inc, Natick, MA, USA) software with the k-Wave ultrasound toolbox~\cite{treeby_k-wave:_2010}.
The tissue medium was discretized on a grid with 75\,$\mu$m lateral and axial resolution.
A circular inclusion with a SoS of 1540\,m/s and a diameter of 6\,mm was modeled in a background substrate with a SoS of 1515\,m/s. 
Speckle echoes were generated by randomly varying the medium density for 40\% of the substrate points with a variation of 20\% around the mean value of 1000\,$kg/m^3$.

The tissue phantom and in vivo experiments were performed using a UF-760AG ultrasound system (Fukuda Denshi, Tokyo, Japan) with 64 Tx and Rx channels, using a FUT-LA385-12P linear array transducer with $N_c$$=$$128$ elements and 300\,$\mu$m pitch.
Data from all 128 transducer elements were collected within two Tx-Rx cycles by repeating the same Tx pulse for storing the Rx echoes consecutively from the left and the right halves of the transducer aperture.
For each Tx, a four-half-cycle pulse of $f_c$$=$$5$\,MHz center frequency is transmitted, followed by RF data reception. 
The received data is stored temporarily in element-wise buffers during the acquisition time from the deepest imaged location, and then transported over a high-speed data-link to an attached PC for storage, before the next Tx-Rx cycle is performed.
This buffer transport leads to an additional overhead of 37.5\,ms between two consecutive Tx events, and is required for downstream SoS processing on PC since we have not implemented SoS reconstruction on US hardware in this work.
This, nevertheless, can be avoided in the future by transporting only the tracked displacements or conducting the subsequent reconstruction on hardware, \eg via~\cite{vishnevskiy2018im} that takes $\approx20$\,ms for inference of an SoS image.

For phantom experiments, a CIRS SoS phantom (Norfolk, VA, USA) was used with a background SoS of 1509\,m/s and 10\,mm-diameter cylindrical inclusions of 1465\,m/s (-3.0\% contrast) and 1585\,m/s (+5.0\% contrast) centered at a depth of 15\,mm from the surface.
In vivo RF data was collected in a clinical study at Kantonsspital Baden, Switzerland, with ethics approval (EKNZ, Switzerland, BASEC 2020-01962), external monitoring and informed patient consent.

\subsection{Evaluation metrics}
To assess accuracy in phantom experiments, we used the root mean square error RMSE=$\sqrt{\frac{1}{N}\sum(\hat{\boldsymbol{c}}-\boldsymbol{c}^\star)^2}$\,, where $\boldsymbol{c}^\star$\, is the ground-truth SoS map, $\hat{\boldsymbol{c}}$ is the reconstructed SoS map, and $N$ the total number of pixels in the reconstructed map. The sum is done over all pixels.
For the ground-truth map, the manufacturer-reported phantom values were used with the inclusion delineated as visible in the B-mode image.
For evaluating contrast in both phantom and in vivo experiments, we used a metric proposed and validated in \cite{Schweizer_SoSasPotentialBiomarker_2025} as $\Delta$SoS = SoS$_\mathrm{Inc}$ - SoS$_\mathrm{BG}$, where SoS$_\mathrm{Inc}$ is the 95th percentile of SoS within the inclusion and SoS$_\mathrm{BG}$ denotes the median background SoS.
The in vivo lesion masks were annotated by a radiologist on the B-mode images. The background region was considered as the rest of the image beyond a 5\,mm margin of the lesion mask.
%%%%%%%%%%%%%%%%%%%%%%%%%%%%%%%%%%%%%%%%%%
\section{Results}

\subsection{Simulation experiments}
\begin{figure*}
\centering
\includegraphics[width=0.9\linewidth]{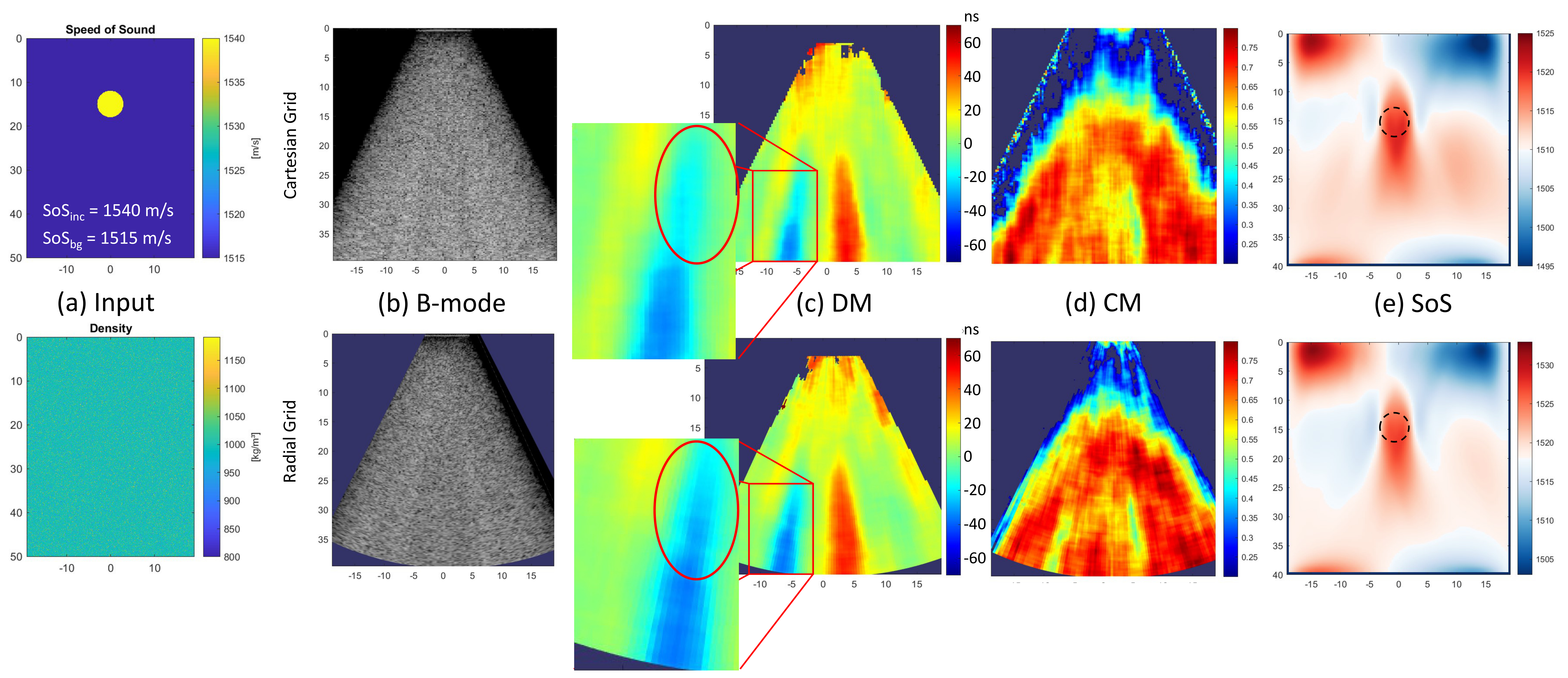} 
\caption[Simulation phantom reconstruction steps]{(a) K-wave phantom, (b) B-mode images for a central Tx pair, (c) Displacement maps (DM), (d) Displacement tracking correlation map (CM), and (e) SoS reconstructions with the inclusion locations overlaid.
Top and bottom rows (b-e) show, respectively, the Cartesian C-grid and radial R$_9$-grid beamforming and their respective postprocessing.
The zoomed-in images show a comparison of measured displacements, with the ellipse indicating an artifactual smoothing and loss in the vertically-tracked C-grid.
Color ranges were offset by $\approx$8\,m/s to adjust for a slight overall SoS difference between the two SoS images to help better compare SoS contrast differences visually.
}
\label{fig:KwaveResults}
\end{figure*}

Figure \Ref{fig:KwaveResults} compares the Cartesian- and radial-grid approaches on the numerical phantom in terms of displacement maps (DM) of $\boldsymbol{\Delta\tau}$, tracked between a sample Tx-pair, which demonstrates a whisker pattern (differential displacement ``shadows'' emanating downward) as Tx travel speeds differ through the inclusion near its edges.
Note that most window-based displacement tracking algorithms lead to an implicit smoothing effect in the tracked direction due to their finite window length. 
As the displacement shadows naturally orient towards the Tx-pair (mid-point), it is also ideal to have displacement tracking aligned in this direction.
It is seen in the zoomed-in windows that C-grid with a vertical beamforming and tracking direction may smooth out and lose subtle displacement cues.
This is further seen in the corresponding cross-correlation maps (CM) indicating unreliable tracking locations from the displacement tracking algorithm.
Although this not visibly observed in the sample SoS reconstruction displayed (due to several pairs combined and regularization smoothed), the loss of such subtle information may affect the visibility of smaller or ambiguous SoS inclusions in other settings.

\subsection{Phantom experiments}
\begin{figure*}
\centering
\includegraphics[width=0.9\linewidth]{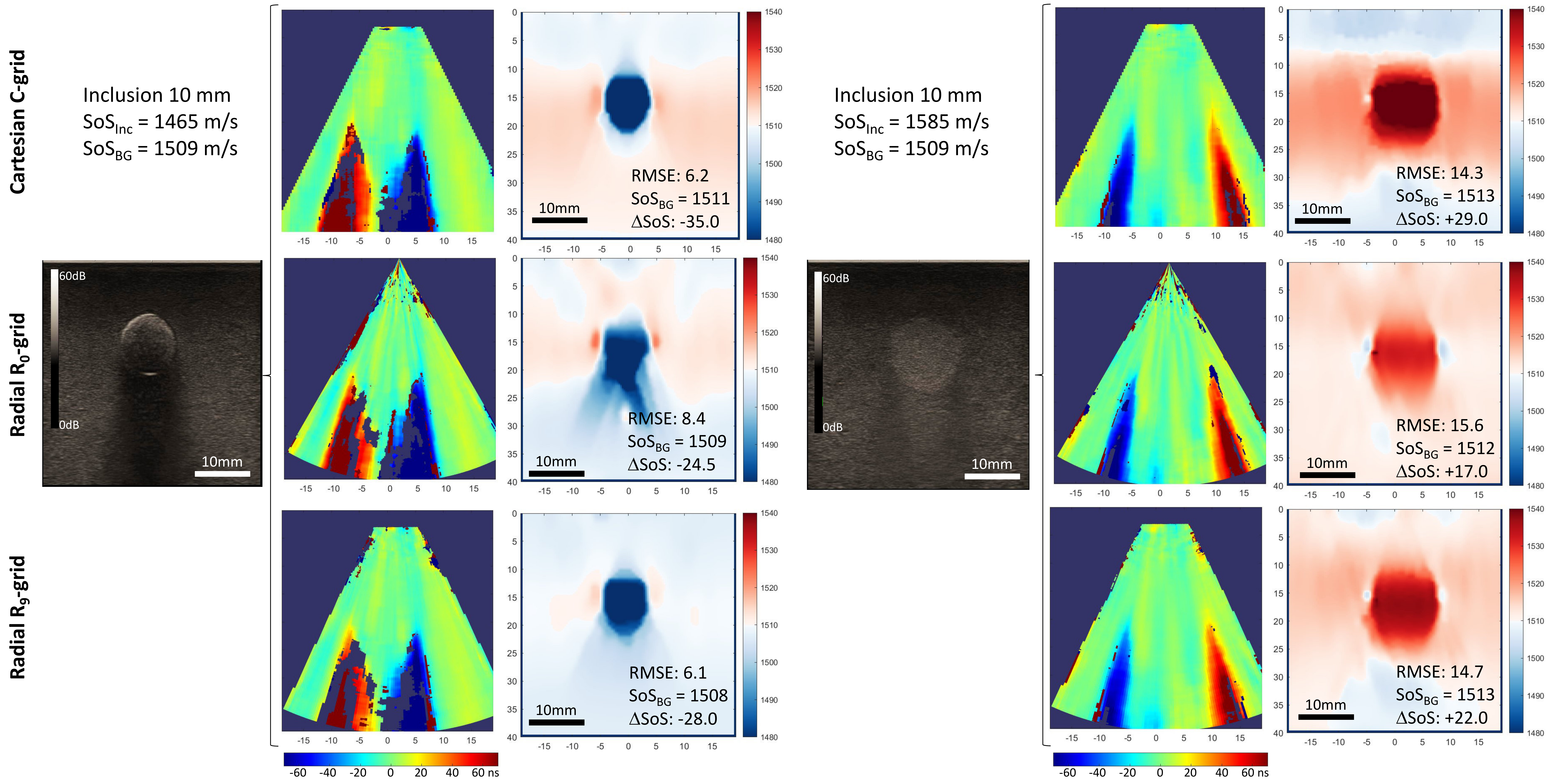} 
\caption[Phantom SoS reconstruction examples]{DMs and SoS reconstructions for two phantom inclusions, with negative (left) and positive (right) SoS contrast. The beamforming method is Cartesian C-grid (top row), radial R$_0$-grid (mid row), and radial R$_9$-grid (bottom row).
RMSE results are given in [m/s].
In C-grid, the DMs are masked with the Tx aperture.
In R-grid, the visible DMs contain all available BF-grid locations, \ie no explicit masking was needed nor applied.
}
\label{fig:TissuePhantomResults}
\end{figure*}

Figure \Ref{fig:TissuePhantomResults} shows DMs and SoS reconstructions with the beamforming approaches.
Displacement tracking artifacts (seen as missing low-correlation values) and hence any window-based smoothing are seen to be better aligned with the observed whiskers for the R$_9$-grid.
Both R-grid solutions show DMs similar to that of their corresponding C-grid alternative. 
While the SoS reconstructions with R$_9$-grid are comparable to that of C-grid, R$_0$-grid is seen to perform worse SoS reconstructions.
This is likely due to the smaller FoV of R$_0$-grid combined with its tracking not best aligning with the Tx directions.
C-grid reconstruction shows moderately higher SoS contrast, but the background SoS is less accurate than that found with the radial grid reconstruction.

\begin{figure}
\centering
\includegraphics[width=\linewidth]{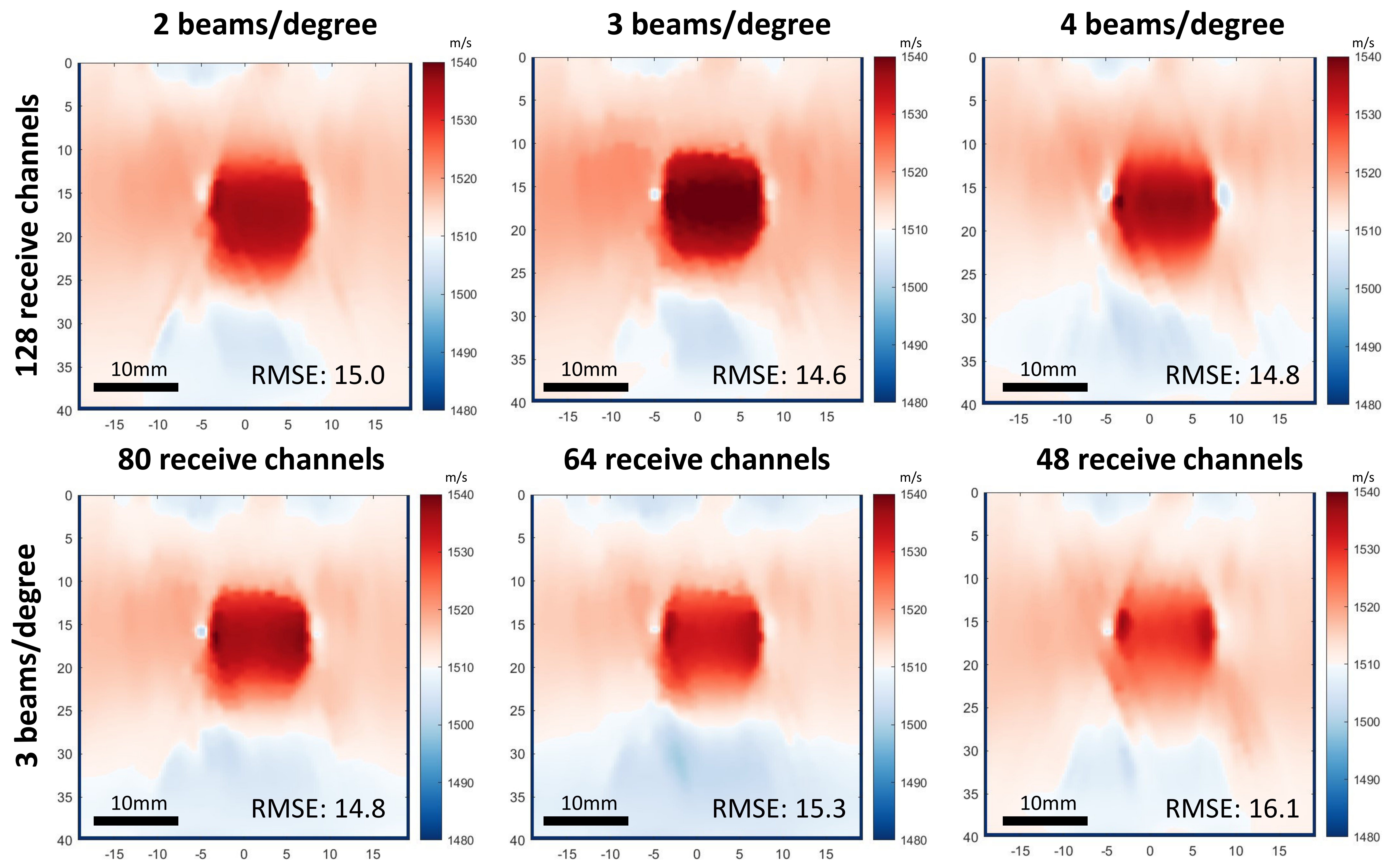} 
\caption[SoS reconstruction with different grid densities]{Radial R9-grid SoS reconstruction of SoS tissue phantoms with radial beamforming: Top row: 128 Rx channels with varying angular densities of beamforming lines.
Bottom row: 3\,beams-per-degree line density with varying numbers of Rx channels.
}
\label{fig:TissuePhantomResultAngleDensity}
\end{figure}

To assess accuracy with different hardware capabilities, Figure \Ref{fig:TissuePhantomResultAngleDensity} compares reconstructions with the radial R$_9$-grid for three different angular resolutions $r$ and for different numbers of maximum Rx channels addressable during beamforming.
The results appear to be comparable, with visual degradation mainly for the 48 Rx channel setting with 4 beams/degree.
Reconstructions in this work have been performed with $r=3$ beams/degree and 128 receive channels.

\subsection{In vivo reconstruction results}
\begin{figure}
\centering
\includegraphics[width=\linewidth]{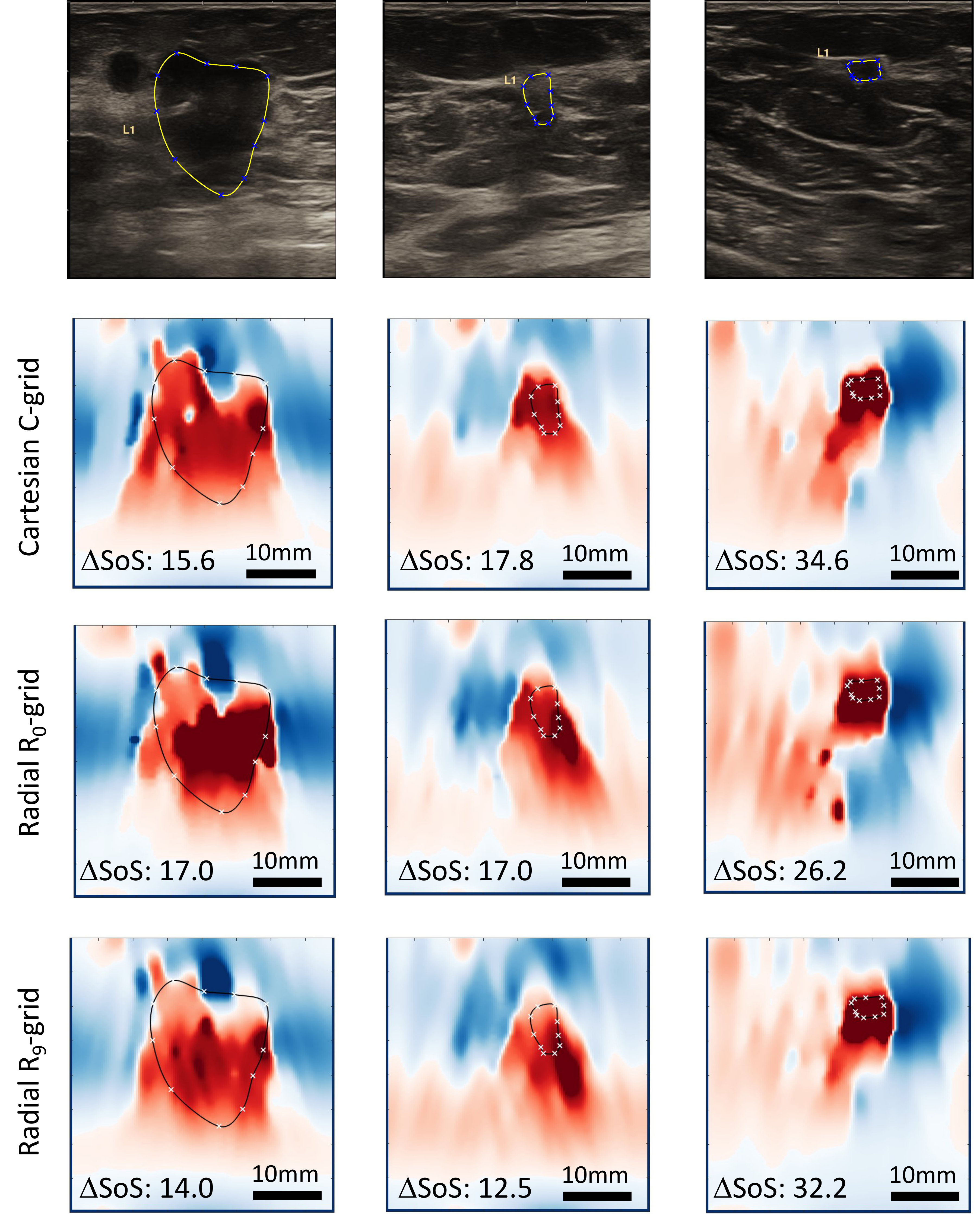} 
\caption[In vivo SoS reconstruction examples]{SoS reconstruction for three breast lesions (ductal carcinomas) of different sizes. 
The reference B-mode image shows the tumor delineation.
The beamforming method is Cartesian C-grid (top row), radial R$_0$-grid (mid row), and radial R$_9$-grid (bottom row). The used color range for each reconstruction within a lesion is the same, corresponding to SoS$_\mathrm{BG}$ $\pm$20\,m/s. $\Delta$SoS is reported in [m/s].
}
\label{fig:InvivoResults}
\end{figure}

\begin{figure*}
\centering
\includegraphics[width=0.6\linewidth]{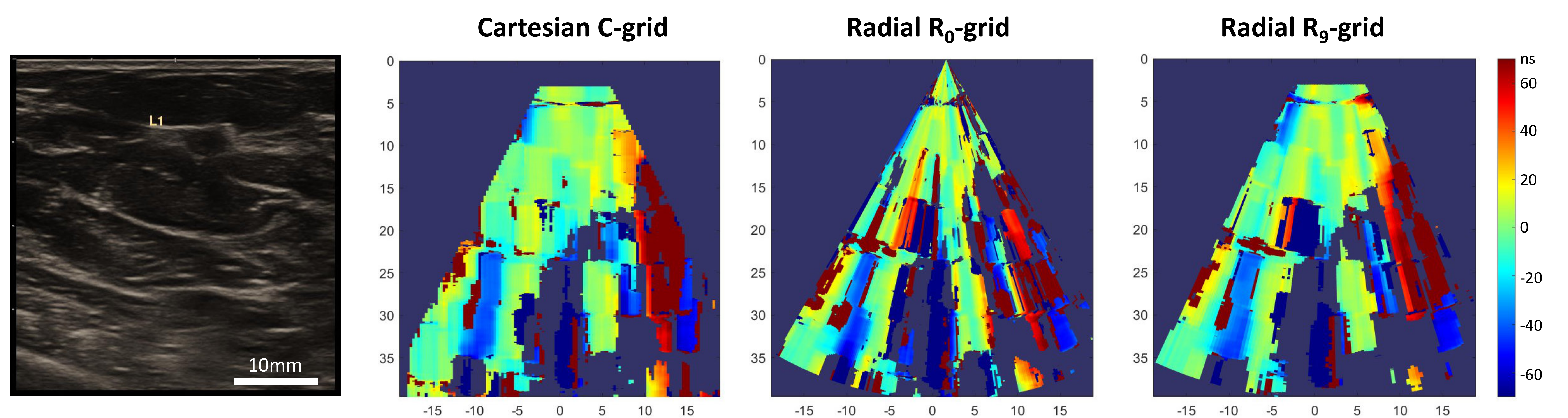} 
\caption[In vivo displacement results]{DMs for a ductal carcinoma shown in Figure \Ref{fig:InvivoResults} (right column) for different beamforming methods.
}
\label{fig:InvivoDTResults}
\end{figure*}

SoS reconstructions from in vivo data of three breast lesions, which are all biopsy-confirmed as invasive ductal carcinoma, are shown in Figure \Ref{fig:InvivoResults}.
Both grid approaches present an increased SoS area inside the lesion, while C-grid shows a slightly higher contrast on average.
Figure \Ref{fig:InvivoDTResults} shows the displacement map data for one Tx-pair of one breast lesion.
Although all DMs contain artifacts due to the complex inhomogeneities of the in vivo setting, R$_9$-grid yields a DM superior to that of R$_0$-grid and comparable to that of C-grid based beamforming.

%%%%%%%%%%%%%%%%%%%%%%%%%%%%%%%%%%%%%%%%%%
\section{Discussion and Conclusions}

In the context of SoS imaging, we have introduced radial beamforming applicable with existing beamformers on low- and mid-range ultrasound systems.
We have presented two variants of this approach, comparing them to the Cartesian-grid beamforming used in~\cite{jaeger_computed_2015,sanabria_spatial_2018,Stahli_improved_20,Rau_divergingWave_2021,Schweizer_RobustVS}.
Our experiments show that with our proposed approach, SoS imaging data can be acquired with over 20\,fps without compromising accuracy compared to software beamforming on a Cartesian grid, which is achievable only on select hardware, \ie advanced, research-grade US machines. 
The radial grid is easier to implement in the front-end, better aligned with the integral displacements for subsequent window-based tracking, and yields PSFs symmetric between the Tx pair.
Among the two radial grid variants we compared, the R$_9$-grid with a beamforming center between the Tx locations allows for an optimal coverage of the valid disparity region and a better alignment of tracking windows with the displacement tracking regions to minimize artifactual smoothing of displacements.
Radial beamforming is not only a step towards real-time SoS imaging but also allows for a natural extension of SoS imaging to trapezoidal image formats, as used for curved arrays.

Fast pair-alternating beamforming (PAB) reduces the time elapsed between beamforming the same point in different frames to a single Tx–Rx cycle, thereby minimizing motion sensitivity to the physical limits of echo ultrasound for consecutive acquisition.
This supports motion-robust acquisition for SoS imaging in practical settings, aiming to enhance SoS reconstruction quality in the presence of patient-intrinsic or operator-induced motion, which are unavoidable and often detrimental to imaging~\cite{Schweizer_RobustVS}.
On-the-fly beamforming methods proposed in this paper can also be implemented in 2D-array front ends used in 3D US systems, since no channel-wise data storage is necessary and pre-beamforming can be done as for typical 3D B-mode imaging.
This opens up the possibility of 3D SoS imaging.
Future work can further optimize Tx-pair separation, f-number setting, and optimal grid resolutions for performance trade-offs between accuracy and computational cost.

\section*{Acknowledgment}
The authors would like to thank the Kantonsspital Baden personnel for the help and support, in particular Prof. Dr. Rahel A. Kubik-Huch, Dr. Monika Farkas, Dr. Anna Potempa, and the study nurse Silke Callies.

\bibliographystyle{IEEEtran}
\bibliography{arxiv}

\end{document}